\documentclass{article}
\usepackage[T1]{fontenc} % add special characters (e.g., umlaute)
\usepackage[utf8]{inputenc} % set utf-8 as default input encoding
\usepackage{ismir,amsmath,cite,url}
\usepackage{graphicx}
\usepackage{multirow}
\usepackage{color}
\usepackage{xcolor}
\usepackage[bookmarks=false]{hyperref}

\usepackage{lineno}
\newcommand{\changes}[1]{#1}
\title{Generating Coherent Drum Accompaniment with Fills and Improvisations}

\multauthor
{Rishabh Dahale$^1$ \hspace{1cm} Vaibhav Talwadker$^1$ \hspace{1cm} Preeti Rao$^1$ \hspace{1cm} Prateek Verma $^2$} { %  \bfseries{Prateek Verma$^2$}\\
 $^1$ Department of Electrical Engineering, Indian Institute of Technology Bombay, India\\
$^2$  Stanford University\\
{\tt\small dahalerishabh1@iitb.ac.in, talwadkerv@gmail.com, prao@ee.iitb.ac.in, prateekv@stanford.edu} }

\def\authorname{Rishabh Dahale, Vaibhav Talwadker, Preeti Rao and Prateek Verma}

\begin{document}

\maketitle
\begin{abstract}
% The abstract should be placed in the top left column and should contain about 150-200 words.

\textbf{Creating a complex work of art like music necessitates profound creativity. With recent advancements in deep learning and powerful models such as transformers, there has been huge progress in automatic music generation. In an accompaniment generation context, creating a coherent drum pattern with apposite fills and improvisations at proper locations in a song is a challenging task even for an experienced drummer. Drum beats tend to follow a repetitive pattern through stanzas with fills/improvisation at section boundaries. In this work, we tackle the task of drum pattern generation conditioned on the accompanying music played by four melodic instruments – Piano, Guitar, Bass, and Strings. We use the transformer sequence to sequence model to generate a basic drum pattern conditioned on the melodic accompaniment to find that improvisation is largely absent, attributed possibly to its expectedly relatively low representation in the training data. We propose a novelty function to capture the extent of improvisation in a bar relative to its neighbors. We train a model to predict improvisation locations from the melodic accompaniment tracks. Finally, we use a novel BERT-inspired in-filling architecture, to learn the structure of both the drums and melody to in-fill elements of improvised music.}

\end{abstract}
\vspace{-5pt}
\section{Introduction}\label{sec:introduction}

Songs in popular music genres like rock are typically split into different sections such as the verse, bridge, and chorus. While the primary task of a drummer is to play in time, it is also important for the drummer to be consistent with the song structure. %add fills that make the piece sound more musically interesting. 
Traditionally fills, or short groups of notes, are played as the song transitions from one section to another (say, verse to chorus). Thus, it can serve as an indicator to the audience as well as the band, of an upcoming transition in the song. The duration of fills generally tends to be only a few beats long, no more than the length of a bar. Even though they are rare events in a drum track, fills are an important part of the overall aesthetics. %as they provide a smooth passage between a drum pattern observed in the verse and the drum pattern observed in the chorus. 
Beyond signaling transitions, drum fills can also be played in sections where the accompanying instrumentation is sparse. %However, the challenging task of these fills would be to engage the listener without overshadowing the melodic accompaniment.

Motivated by the above, we improve the quality of the generated drum tracks from seq-to-seq models by incorporating fills/improvisations towards our overall goal of accompaniment generation. This is achieved via the following three distinct stages:

1. \textbf{Basic Drum Pattern Generation:} Using a seq-to-seq model to generate a drumbeat as the accompaniment to given melodic instrument tracks of guitar, bass, strings, and piano (i.e. the Melodic Accompaniment - MA).

2. \textbf{Improvisation Location Detection:} Detecting explicitly the position of improvisation from the MA using a self-similarity function and mini BERT model.

3. \textbf{Generating Improvised Bars:} Generating the fills in the previously detected bars.

Our main contributions are:
(i) We show that traditional attention-based transformer architectures fail to capture the ``improvisation" due to implicit data imbalance.
(ii) We also show that the sampling-based approaches fail to produce a variation of pattern at the right location. To mitigate this, we learn to predict where to improvise directly from the melody tracks using powerful self-attention-based architectures.
(iii) We propose a novel in-filling approach, inspired by BERT that can look at the context of drums and the context of melody and use it to generate the improvised bars.
(iv) We demonstrate an MLP-based synthesis module for drum improvisation generation from a latent code.
% \changes{In this work, we don't focus on the dynamics (velocities) of the generated drum beat; instead, we concentrate on generating good drum patterns with fills.}
\changes{For simplicity we have ignored the dynamic (velocities) in the generated drum patterns of this work.
% In this work, we focus on generating good drum patterns with fills and not on the dynamics (velocities).
}
%(v) This work shows a serious drawback of traditional language-based generators, which have shown promise in a lot of different fields, yet they fail to capture subtle musical signals, where they are often sparsely occurring in otherwise repetitive and common patterns.

\vspace{-10pt}
\section{Related Works}\label{sec:related_works}

% From Hidden Markov Models to RNNs and Long Short Term Memory networks, sequence models have been the standard choice for modeling music (Eck at al. \cite{eck2002finding}, Liang et al. \cite{liang2016bachbot}, Oore et al. \cite{oore2020time}). With the advent of transformer architecture by Vaswani et al. \cite{vaswani2017attention}, many sequence modeling jobs have been transferred to transformers.

% Conditioned drum pattern generation, a critical subtask of music generation, necessitates the generation of rhythmic patterns.

\begin{figure*}[t]
    \centering
    \includegraphics[width=0.9\linewidth]{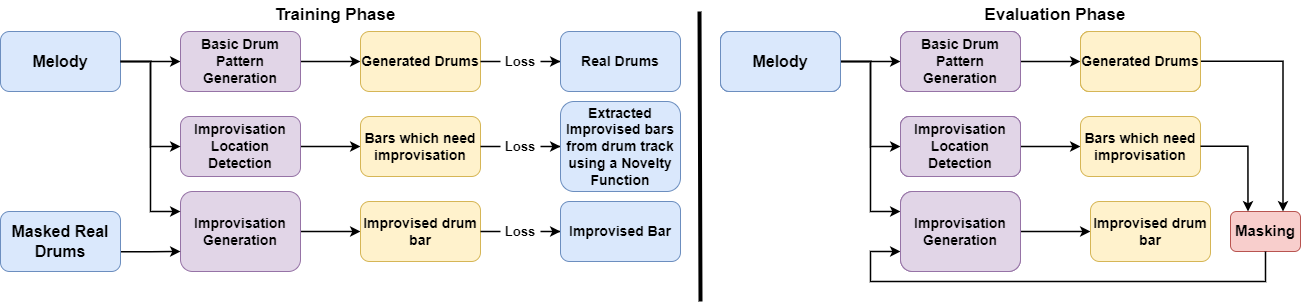}
    \vspace{-10pt}
    \caption{System Overview - Pipeline used for training (left half) of each individual module; combined pipeline (right half) for the evaluation phase.}
    \label{fig:system_overview}
\end{figure*}

Since the introduction of the Transformer architecture \cite{vaswani2017attention}, there has been a growing interest in this model for sequential task modeling like in NLP and music generation. The architecture uses an attention mechanism to learn long-term patterns and can easily surpass dilated convolutional-based methods such as WaveNet \cite{verma2021generative}.
% The Transformer architecture learns long-term relationships in its inputs through attention mechanisms, allowing it to learn long-term patterns, surpassing dilated convolutional-based methods such as wavenet \cite{verma2021generative}.
They achieve state-of-the-art performance in a variety of natural language problems \cite{brown2020language}, music/audio understanding \cite{verma2021audio} and generation tasks \cite{dhariwal2020jukebox, huang2018music}. However, for generative models, the decoding strategy still remains an open question. Even though high-quality models can be obtained by the use of likelihood as the training objective, likelihood maximization methods like beam search lead to degeneration \cite{holtzman2019curious}. To solve this issue, many researchers use sampling-based methods like temperature sampling, top-k sampling, and nucleus sampling \cite{ficler2017controlling, fan2018hierarchical, holtzman2019curious}. Despite all of the success of sequential modeling recently, there still exist many issues that are relevant to our current work such as understanding rare words or sparsely occurring events of interest \cite{schick2020rare}. Another major problem is the presence of biases in the generated output, as they mimic the distribution present in the training datasets \cite{sheng2019woman}. These issues are ubiquitous across datasets and modalities and are implicit in our task due to the low representation of improvised bars. We here show how these constraints and biases affect the quality of drum generation, and the steps we take to mitigate them. %The work here can go beyond the problem of interest, where similar issues are present.
% Huang et al. \cite{huang2018music} demonstrated compelling results by using transformers to generate long-duration piano music. They propose an algorithm that reduces the intermediate memory requirement to be linear in sequence length as opposed to quadratic in sequence length as proposed by Vaswani et al. \cite{vaswani2017attention}.

Conditioned drum beat generation is an important subtask of music generation. Wei et al. \cite{wei2019generating} observed that the polyphonic melody self-similarity matrix (SSM) is structurally similar to the drum SSM. They used this to predict the drum SSM from the melody SSM and used this intermediate representation to generate the drumbeat. While they used the audio form of the melody as the input, the symbolic domain also offers opportunities for similar research. \changes{An example is the work of \cite{tamagnan2019drum} using symbolic representation of the drum track to predict locations and generate improvisations for the drum track. In this work, we address drum generation conditioned on a melodic accompaniment track, bringing considerably more complexity to the problem.
% This is not a good method as improvisations in the drum track are highly motivated by the melodic accompaniments. Further, their work used VAE features, to classify if a single drum bar contains improvisations or not. We in this work, use a BERT inspired architecture to learn where improvisation is needed directly from melodic features, and use a separate architecture using melody and generated neighbouring drum pattern to synthesize those fills.
% Additionally, we use  a BERT inspired approach that can synthesize the fills using the left context, the right context and the entire melody to learn a latent representation used by fill synthesis module.
}

In the symbolic domain, systems use a discretized representation such as MIDI tag and pianoroll representations that capture the essential information at the semantic level. Pianoroll is a score-like matrix representing a piece of music. Note pitch and time are represented by the vertical and horizontal axes, respectively. The velocities of the notes are represented by the values. The time is generally quantized on a sub-beat level and each instrument track is represented by a separate pianoroll matrix. Dong et al. \cite{dong2018musegan} used this representation along with CNN-based GAN model for music generation. Another popular symbolic domain representation is the MIDI tag representation which we refer to as the ``serialized grid representation".  In this method, the input is represented by a sequence of MIDI-like tags. Huang et al. \cite{huang2018music} used this MIDI tag representation with modified transformer architecture to generate long-duration piano music. Several modifications have also been proposed to this representation. Huang et al. \cite{huang2020pop} proposed a revamped MIDI (REMI) which introduced \texttt{DURATION}, \texttt{BAR} and \texttt{POSITION} tags to improve the quality of generated music. Ren et al. \cite{ren2020popmag} further modified this representation to include multi-track representation by introducing \texttt{TRACK} tag and used the Transformer-XL model to generate multitrack songs.
% Several modifications have been proposed to this representation. Huang et al. \cite{huang2020pop} proposed a revamped MIDI (REMI) which introduced \texttt{DURATION}, \texttt{BAR} and \texttt{POSITION} tags. Ren et al. \cite{ren2020popmag} further modified this representation to include multi-track representation by introducing \texttt{TRACK} tag.
% Dong et al. \cite{dong2018musegan} used a CNN-based GAN model to imitate the music generation process with piano-roll representation using an inter-track latent vector to generate a coherent output. 
Nuttall et al. \cite{nuttall2021transformer} modified the MIDI tags of nine percussion instruments to represent notes being played by a triplet of pitch, velocity, and start time, and used it with the Transformer-XL model to sequentially generate the drum pattern. Thorn et al. \cite{thorn2020ai} demonstrated three experiments with the Transformer-XL model with varying input and output representation and control. In two of the experiments, they tried to control the drum machine while in the third experiment they tried to generate the drum pattern directly. %We do not use Transformer-XL, or any of the subsequent attention based advanced architectures, as they all suffer from the same issues of biases in dealing with the implicit data-imbalance that exists in the training dataset \cite{caliskan2017semantics,huang2019reducing}.  
% While the authors present results that are not much meaningful concerning the drumming, we find that the Transformer model proposed by Vaswani et al. \cite{vaswani2017attention} was able to generate a good basic drum pattern.

While the Transformer-XL model facilitates the generation of longer duration (musical) sequences, they still suffer from the same issues of biases in dealing with the implicit data-imbalance that exists in the training dataset \cite{caliskan2017semantics, huang2019reducing}. 
As the specific focus of this work is to find ways to capture the rare events in the generated output, i.e., fills and improvisations, we consider only the necessary context for an improvised bar in the form of 11 bar segments of the songs. %This also helps us limit the needed computing resources. %while not available to us. We find the Transformer model proposed by Vaswani et al. \cite{vaswani2017attention} is able to generate a good basic drum pattern for this task.

\vspace{-10pt}
\section{Dataset}\label{sec:dataset}
In this work, we use the Lakh Pianoroll Dataset (LPD-5 cleansed) \cite{dong2018musegan}, derived from the Lakh Midi Dataset \cite{raffel2016learning} which is a collection of 21,425 multitrack pianorolls files consisting of following tracks: Piano, Guitar, Bass, Strings and Percussion. All the songs in this dataset are of 4/4-time signature i.e., all the songs contain 4 beats in a bar and the dimensionality of each bar in this dataset is 128 (pitch) x 96 (time steps) i.e. each beat is divided into 24 parts. \changes{} Our input, which we call melodic accompaniment (MA), consists of notes played by the 4 melodic instruments and output is a percussion instrument pattern conditioned on the input which we call the percussion accompaniment (PA).

Evaluation of generative music systems faces harder challenges than that of image generation systems \cite{briot2017deep}. We expect our system to replicate the rhythmic consistency and diversity of the dataset. Any drum beat generation system must have correct onset locations in a beat. If the onsets are not properly matched, it appears as if the drums are lagging/leading the melody.
% The most basic requirement for any drum beat generation system is to have accurate onset locations in a bar. If the onsets are not properly aligned with the beats, it gets perceived as if the drums are lagging/leading the melody.
% a consistent onset positioning.
We show that our model is able learn this by capturing the onset location distribution. To compare generated samples with original drums in the dataset, we use the following metric: Instrument Count - Number of distinct percussion instruments used in a bar. To evaluate our outputs in terms of rhythmic consistency, we use the following objective metric: Percussion pattern consistency in consecutive bars.
% We use instrument density and stroke density to compare the diversity present in generated outputs with the diversity present in the dataset and use three more objective metrics to show that the generated outputs have a rhythmic consistency. We also present some generated samples to trained musicians and report their comments. 
% We use objective evaluation measures to compare the diversity and rhythmic consistency of the dataset with the generated samples.

\vspace{-10pt}
\section{Method}\label{sec:method}

The overview of the proposed system is shown in \figref{fig:system_overview}. Details of each of the models are provided in subsequent sections. Our models are trained for 300 epochs with Adam optimization \cite{kingma2014adam} starting with a learning rate of 1e-4 and decaying it till 1e-6. All the setup was carried out using the Tensorflow \cite{abadi2016tensorflow} framework. The following two modules are being used in all the models:\\
1. \textbf{Embedding Module:} As the pianoroll matrices are highly sparse, we pass them through 2 layers with 1024 and 128 ReLU \cite{agarap2018deep} activated dense layers to capture the inter instrument and inter pitch dependencies. For the decoder branch of Basic Drum Pattern Generation model, we use 128 dimensional token embedding layer.\\
2. \textbf{Position Encoder:} We concatenate the sinusoidal positional representations \cite{vaswani2017attention} with the 128 dimensional vectors and use a dense layer to project them back in 128 dimension space.
% The sections below contain information about each individual model:

% The overview of the proposed system is shown in Figure \ref{fig:system_overview}. The generation is split into 3 parts: (i) Basic Drum Pattern Generation which generated an 11 bar drum pattern for the 11-bar of melody. We restrict ourselves to 11 bars as the model for generating an improvised bar takes the bar which needs to be improvised along with a context of 5 bars on both sides for both melodic accompaniment and percussion accompaniment; (ii) Detection of bars that need improvisation and; (iii) Generating improvised drum bars for the locations detected. These improvised bars generated by the final model are then replaced by the basic bars in the percussion track generated by the first model. Data representation and model details are given in the following subsections.

\vspace{-5pt}
\subsection{Data Processing \& Representation}\label{subsec:data_processing_repr}

To compress the input and output representation in our work, we perform the following preprocessing steps:
(i) trim the track for start and end silence bars;
(ii) resample the beats to 8 parts per beat, making each bar 32 timesteps long;
(iii) keep only active MIDI pitches in all melodic instruments, i.e. MIDI 21 to MIDI 83 (notes A0 to B5);
(iv) combine similar instruments in the MIDI representation of the percussion track. For example, under the snare drum, MIDI 38, which corresponds to acoustic snare, and MIDI 40, which corresponds to electric snare, are clubbed together;
(v) choose only 16 percussion instruments as they capture 85.3\% of all the percussion instrument strokes: snare drum, open hi-hat, close hi-hat, kick drum, ride cymbal, crash cymbal, low-floor tom, high-floor tom, high tom, hi-mid tom, low tom, cowbell, pedal hi-hat, tambourine, cabasa, and maracas;
(vi) to decrease the amount of training parameters, we binarize the percussion track for seq-2-seq models and exclude velocities;
(vii) split the song in non-overlapping contiguous 11 bar samples.
% from the sequence length.
% (v) binarize the percussion track for seq-2-seq models and omit velocities to shorten the sequence length, due to limited availability of computing resources. They will only have a marginal effect on the final output, as a signal of presence/absence of drum instrument is more salient in determining what to play next.
\changes{The finer grids are superior for representing drum audio and fills\cite{gillick2021drumroll}.
% We however opted for quantized representation, as it was advantageous for our powerful architectures, both for understanding and synthesis, to learn dependencies.
We however opt for a quantized representation, capturing most of the significant musical events, allowing us to model longer duration dependencies by the compressed representation.
It will be interesting to compare various representations as a future work.
}

\begin{figure}[ht]
    \centering
    \includegraphics[width=\linewidth]{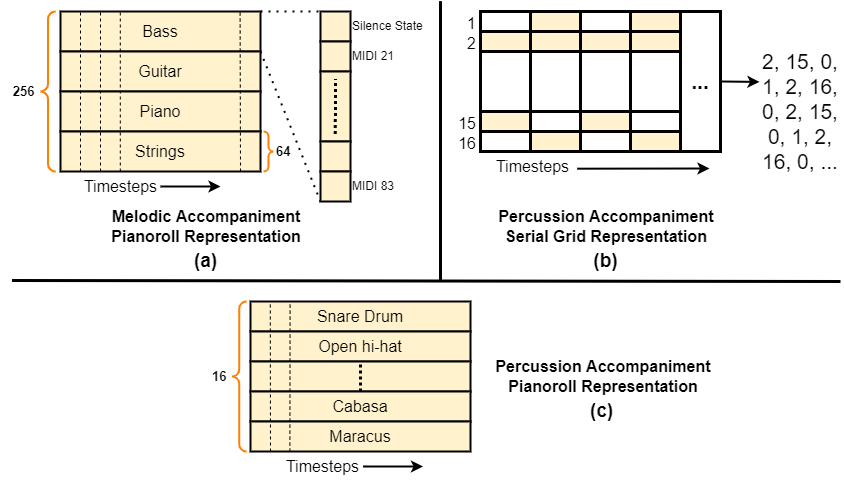}
    \caption{Data representation methods used in this work% (a) Melodic accompaniment pianoroll representation; (b) Percussion accompaniment serial grid representation; (c) Percussion track pianoroll representation
    }
    \label{fig:data_repr}
    \vspace{-10pt}
\end{figure}

We use a modified version of pianoroll representation for MA representation. We concatenate the pianorolls (\figref{fig:data_repr}a) of different instruments instead of adding them to different channels \cite{dong2018musegan}. As our input is quantized to 8 parts per beat, we represent each $ \frac{1}{8}^{th} $ part of the beat by a 256-dimensional vector split  amongst the 4 melodic instruments. The first dimension of this 64-dimension vector is the silence state. This is a binary state representing if the instrument under consideration is silent. 
% If this state is 0, the rest of the 63 dimensions contain the velocities of the MIDI notes 21-83 being played.
The rest of the 63 dimensions contain the velocities of the MIDI notes 21-83 being played.

% We modify the piano-roll representation to represent the melodic accompaniment. Instead of the popular method of using separate channels to represent different instruments (Dong, et al. \cite{dong2018musegan}) we concatenate these matrices as shown in figure \ref{fig:data_repr}. As our input is quantized to 8 parts per beat, we represent each $ \frac{1}{8}^{th} $ part of the beat by a 256-dimensional vector which is split equally amongst the 4 melodic instruments (i.e. 64 dimensions for each instrument). The first dimension of this 64-dimension vector is the silence state. This is a binary state representing if the instrument under consideration is silent. If this state is 0, the rest of the 63 dimensions contain the velocities of the MIDI notes 21-83 being played.
For the PA, we adopt a mixed representation. For our Basic Drum Pattern Generation model, which is a transformer seq-2-seq model, we take advantage of the language modeling tasks and use a serialized grid representation (\figref{fig:data_repr}b).
In this representation, only the active percussion instruments which are being played are unfolded into a sequence of tokens.
We add a silence state token and shift by one token making a total of 18 tokens for the percussion track.
For the final model, the improvisation generation model, we give the MA and masked basic PA pattern as the inputs. We use the pianoroll representation for the PA representation for this model (\figref{fig:data_repr}c) as only a fixed number of timesteps needs to be masked in this representation. %With serial grid representation, the number of tokens required for every bar varies according to the density of the strokes in the bar. (Reasons: simple to mask the drum bar which needs to be improvised, generating a fixed shape output (32x16) easier than generating a sequence of tokens with an unknown length which will require asynchronous decoding)

\vspace{-10pt}
\subsection{Train-Test Splits and Data Augmentation}\label{subsec:data_augmentation}

We split the 21,425 songs in LPD-5 into 16,832 songs for training and 4,593 songs for the validation set. As the number of songs is limited, we use the validation set as the test set. We apply the following data augmentation strategies (inspired by sensor dropout methods in robotics \cite{liumulti}) to all of our models' inputs to increase the robustness in the training process:

% To have a more robust model, we deploy the following data augmentation methods (inspired by sensor dropout methods in robotics \cite{liumulti}) to all the inputs of our models:
1. \textbf{Random instrument masking:} We randomly mask one of the instruments in MA for 40\% of the samples in every epoch. This 40\% is equally split between the four melodic instruments. Musically this implies that one of the instruments has stopped playing. This encourages the model to consider all the instruments in the MA while making a prediction.

2. \textbf{Random timestep masking:} We randomly mask 20\% of the timesteps in every sample. Musically this leads to a small disruption in the rhythm of the song which helps in better generalization of the model.

For the improvised bar generation model, which takes in the MA and masked PA as inputs, (section \ref{subsec:improv_generation}), we use the following additional augmentation methods:
% in addition to the ones present above:

1. \textbf{Input masking:} We randomly drop one of the inputs (MA or PA) to the model in 20\% of the input samples. This ensures that the model is not dependent on only one input for improvisation generation.

2. \textbf{Drum noise:} In the evaluation phase, the drum inputs are taken from the output of the basic drum pattern generation model. As this process could be prone to errors, we simulate this by adding the random noise to the drum samples while training. The random noise can either add new drum strokes or remove some old strokes. Our analysis showed that the PA has a very low density in the pianoroll format (average $\approx$ 5\%). Hence we perturb the PA density by a maximum of 1\% % As the drum piano roll has a very less density (average  5\%) We alter the input drum density by a maximum of 1\%

\subsection{Basic Drum Pattern Generation}\label{subsec:baseline_model}

\begin{figure}[ht]
    \centering
    \includegraphics[width=0.9\linewidth]{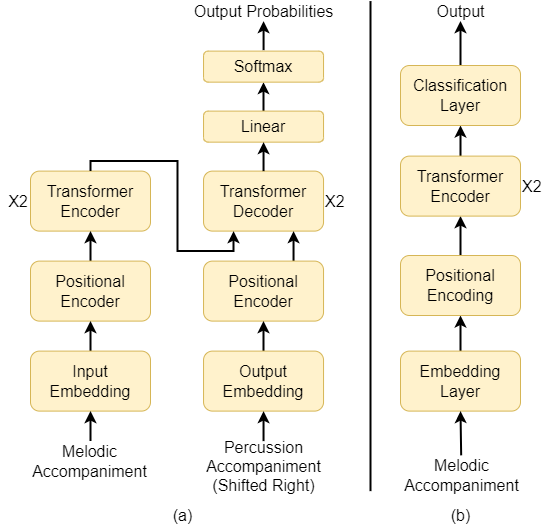}
    \caption{(a) Sequence to sequence model, used for basic drum pattern generation; (b) Improvisation Location Detection Model}
    \label{fig:basic_drum_pattern_generation}
    \vspace{-10pt}
\end{figure}

We use the Transformer encoder-decoder model \cite{vaswani2017attention} to generate a basic drum pattern (\figref{fig:basic_drum_pattern_generation}a). This is done by giving the MA as the input to the encoder and the shifted PA tokens to the decoder branch. Both the inputs are first passed through the embedding module followed by the position encoder. The embedded inputs are passed through 2 layers of encoder/decoder module with 128 dimensional latent space and 8 attention heads. At the output, we have 18 neurons corresponding to the 16 drum instruments, silence token, and shift token.

\subsubsection{Sub-module Evaluation}\label{subsec:baseline_results}

We evaluate the above model with the negative log-likelihood (NLL) values over the train and the validation set. As the model outputs a distribution over the 18 output tokens, there are multiple ways to decode it. We test the greedy method of decoding, where the token with the maximum probability is selected at every step and simple sampling method. The model is trained using categorical cross-entropy loss, achieved a NLL of 0.108 and 0.112 on the train and validation split, respectively.
% The above model is trained using categorical cross-entropy loss and was able to achieve a negative log-likelihood of 0.108 and 0.112 on the train and validation split respectively. We test multiple output decoding methods: greedy search, temperature sampling, top-k sampling, nucleus sampling, and beam search. <EVALUATION OF DECODING METHODS>.

\subsection{Improvisation Location Detection}\label{subsec:improv_detection}

\subsubsection{Novelty Function}\label{subsec:improv_detection_novelty}

\begin{figure}[ht]
    \centering
    \includegraphics[width=\linewidth]{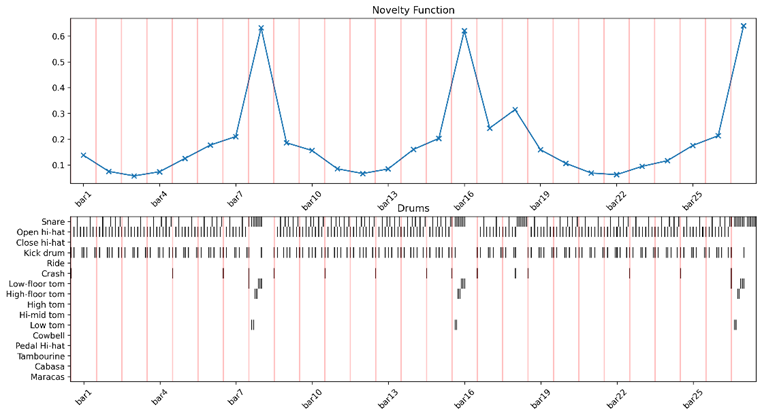}
    \caption{Novelty Function plot for a given drum track}
    \label{fig:novelty_plot}
    \vspace{-10pt}
\end{figure}

% Fills/Improvisations are locations where the drummer makes changes in the basic drum pattern. As these are rare events, they get overshadowed by repeating drum patterns. 
In order to extract locations in the MA that warrant a fill, we propose the following method:

1. We use a 11-bar drum sample to calculate the Novelty value of center bar. The 5 bars on it’s left and right are the context bar. The novelty value of a bar is calculated as the average weighted dissimilarity over all the context bars. We use the following equation to calculate the dissimilarity between 2 bars:
\begin{equation}\label{equ:distance_formula}
    \frac{||bar_i - bar_j||_1\times k}{||bar_i||_1 + ||bar_j||_1}
\end{equation}
We utilize the hanning window to represent the weighting parameter $k$.
    
2. This calculation is done for all the bars across a track, except the first and the last 5 bars due to the lack of context bars. From \figref{fig:novelty_plot} it can be seen that the novelty function peaks at the bar with a drum fill. These bars are then extracted using a peak picking mechanism.
    
3. To generate the dataset for our task, we pick the bars with a local maxima as the positive samples. To filter out minor deviations, e.g., bar 18 in \figref{fig:novelty_plot}, we put a threshold of 0.1 on the peaks height difference from its neighbours. Maximum of 10\% of total bars in a song with these characteristics are selected as the positive samples. Same number of bars from the rest of the non peak regions are selected as negative samples.

\subsubsection{Model Architecture}\label{subsec:improv_detection_model}

% \begin{figure}[ht]
%     \centering
%     \includegraphics[width=0.3\linewidth]{figures/improv_detection_model.png}
%     \caption{Improvisation Detection Model}
%     \label{fig:improv_detection_model}
% \end{figure}

We use a 2 layer BERT \cite{devlin2018bert} (\figref{fig:basic_drum_pattern_generation}b) to detect the location of improvisations. The input to this model is an 11-bar MA and the prediction is done for the middle bar. The MA is passed through an embedding layer followed by the positional encoder. The embedded inputs are then passed through 2 layers of transformer encoder with 64 dimension latent space and 12 attention heads, followed by 1024 neurons. These are finally passed to a 2 dimensional softmax activated dense layer which acts as a classification module. The above model is trained using Huber loss \cite{huber1992robust}, as it is robust to outliers and less sensitive to noise. 

\subsubsection{Sub-module Evaluation}\label{subsec:improv_detection_results}

We monitored the accuracy, precision, and recall of the model in terms of detecting the improvised bars where the target is the original drum track. The final results can be found in Table \ref{tab:result_improv_location}. As the dataset is split equally between positive and negative samples, we have balanced precision and recall values.

% \usepackage{graphicx}
\begin{table}[ht]
\centering
\resizebox{0.43\textwidth}{!}{%
\begin{tabular}{l|c|c|c|c|}
\cline{2-5}
                                 & Precision & Recall  & Accuracy & F1 Score \\ \hline
\multicolumn{1}{|l|}{Train}      & 92.3   & 92.3 & 92.3  & 92.3  \\ \hline
\multicolumn{1}{|l|}{Val} & 79.1   & 79.4 & 79.3  & 79.2  \\ \hline
\end{tabular}%
}
\caption{Performance of the Improvisation Location Detection model. All the values are in \%.}
\label{tab:result_improv_location}
\end{table}

\vspace{-10pt}
\subsection{Improvisation Generation}\label{subsec:improv_generation}

\begin{figure}[ht]
    \centering
    \includegraphics[width=0.8\linewidth]{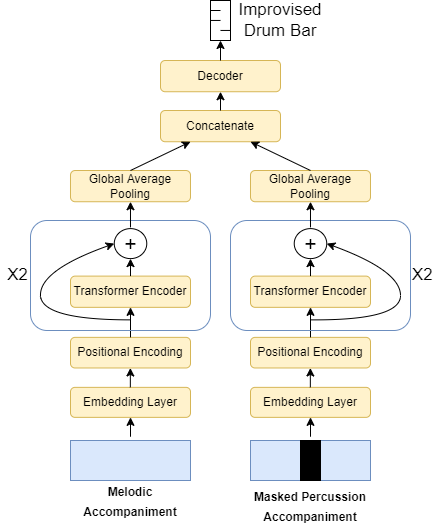}
    \caption{Improvised Bar Generation Model}
    \label{fig:improv_generation_model}
    \vspace{-10pt}
\end{figure}

The final step in our system is the generation of improvised bars. To achieve this we use the architecture shown in \figref{fig:improv_generation_model}. We provide 11 bar MA and PA as the input to the model. During the training phase, the PA is the original percussion track, whereas, during the evaluation/generation phase, the percussion track generated by our first stage model (section \ref{subsec:baseline_model}) is used. The 6th bar in the percussion sample (middle bar) is the target bar and is masked while giving as the input.\\
We generate a summary vector of both the MA and the PA inputs. Both are first passed through an embedding layer followed by a position encoder module. This is then passed through 2 layers of transformer encoders with 128 dimensional latent space and 8 attention heads.
\changes{
% Given more data and bigger network, we expect the infiling detecting and generation system to improve.
Even though larger/bigger models could potentially lead to better results, we have used the resources at our disposal. We note that any further improved performance will only apply to in-fill detection and synthesis.
}
% \changes {While we would certainly get improved scores with bigger/larger models (and there is no end to empirical evaluation), given the resources we had, this is what we could do. However the better performance will only be in terms of in-fill detection and synthesis. The basic drum generation model will still suffer from data-imbalancing no matter how large/bigger models we choose to work with. }

We add skip connections to ease the flow of gradients \cite{pepino2021emotion}. To generate the summary vector for each input, we use a global averaging technique. These two vectors are concatenated and passed through a decoder structure which looks at the concatenated vector to generate the improvised drum bar. We test the following decoder architectures with our model:

1. \textbf{MLP:} A 3-layer dense network with 2048-2048-512 neurons is used. The final layer is sigmoid activated. The outputs are reshaped to 32 (timesteps) $\times$ 16 (percussion instruments) to get the improvised bar.
    
2. \textbf{MLP mixer:} MLP mixers \cite{tolstikhin2021mlp} are simple alternatives to convolution and self-attention. They are based on multi-layered perceptrons applied across either temporal dimension or feature dimension.

3. \textbf{Conv1d:} A simple conv1d architecture with blocks of 2 layers of conv1d followed by upsampling.

\changes{We did not opt for an auto-regressive architecture, as this work does not assume causality, and we incorporate the right context as well as melody for the fill synthesis. There have been other works for improvisation synthesis, e.g. \cite{gillick2019learning}, also using the left and right context, even if only using drums}

% \item Conv2d: We adapted the decoding block from the pix2pix network by Isola et al. \cite{isola2017image}. In this decoding block, instead of using an upsampling block, a stride of 2 is used in the convolutional-transpose filters.

\subsubsection{Sub-module Evaluation}\label{subsec:improv_generation_results}

We treat the prediction of the improvised bars as a regression problem.
% , due to lack of structure present in these bars, compared to other bars.
We similarly train it with Huber loss as it is less sensitive to outliers. We do not use cross-entropy loss for generation firstly purely as a design choice, and intuitively each of the time step token in the prediction within a bar lack probabilistic interpretation. We monitor and report the accuracy, precision, and recall of the models. As the distribution of 0s and 1s is not uniform in the predicted sample, F1 score provides a better insight in the performance of the models. From Table \ref{tab:improv_gen_results} it can be seen that a simple 3 layered MLP decoder is able to perform better than the complex MLP mixer and Conv1D architecture.

% \usepackage{multirow}
% \usepackage{graphicx}
\begin{table}[ht]
\centering
\resizebox{0.48\textwidth}{!}{%
\begin{tabular}{ll|c|c|c|c|}
\cline{3-6}
 &
   &
  \multicolumn{1}{l|}{Precision} &
  \multicolumn{1}{l|}{Recall} &
  \multicolumn{1}{l|}{Accuracy} &
  \multicolumn{1}{l|}{F1 Score} \\ \hline
\multicolumn{1}{|l|}{\multirow{2}{*}{MLP}} &
  Train &
  98.8 &
  93.2 &
  86.3 &
  95.9 \\ \cline{2-6} 
\multicolumn{1}{|l|}{}                        & Val   & 82.9          & \textbf{70.3} & 79.0          & \textbf{76.0} \\ \hline
\multicolumn{1}{|l|}{\multirow{2}{*}{\begin{tabular}[c]{@{}l@{}}MLP\\ Mixer\end{tabular}}} &
  Train &
  97.5 &
  45.0 &
  88.7 &
  61.6 \\ \cline{2-6} 
\multicolumn{1}{|l|}{}                        & Val   & \textbf{83.5} & 42.1          & \textbf{86.0} & 56.0          \\ \hline
\multicolumn{1}{|l|}{\multirow{2}{*}{Conv1D}} & Train & 55.2          & 70.6          & 86.2          & 61.7          \\ \cline{2-6} 
\multicolumn{1}{|l|}{}                        & Val   & 53.1          & 70.3          & 86.1          & 60.5          \\ \hline
\end{tabular}%
}
\caption{Results for various decoders used in the Improvised Bar Generation model (all the values are in \%)}
\label{tab:improv_gen_results}
\end{table}

\vspace{-10pt}
\section{Evaluation}\label{sec:evaluation}

% \subsection{Compared Methods}\label{subsec:evaluation_methods}

To evaluate the quality of the generated PA pattern of the proposed system, we conduct both objective and subjective tests\footnote{Note:Additional objective evaluation methods are reported in the supplementary document \url{https://bit.ly/2022ismirsupp}} with:
\textbf{O:} Original MIDI drum patterns from the dataset;
\textbf{P1:} The basic drum pattern generated by our Basic Drum Pattern Generation model (section \ref{subsec:baseline_model});
\textbf{P2:} The final drum pattern with fills and improvisations generated by the complete system.

% \vspace{-10pt}
% \subsection{Basic Filtration of the Output}\label{subsec:filtering}

% As the output of the Basic Drum Pattern Generation model is decoded using a sampling-based method, it is prone to errors. Sampling a wrong output at any step can lead to degeneration. To reject such samples, we apply the following filters:

% 1. Getting a complete silence after a couple of drum beats was the most common observed output degeneration pattern. To filter out such samples, we check for the total number of silence bars in the generated outputs and reject all the generated patterns with more than 4 silence bars.
% % While short silences are an important part of music, having a very long pause ruins the overall feel. For all the generated samples, we check the number of silence bars and reject samples with 4 or more silence bars.

% 2. While sampling from a distribution, states with lower probability can also be sampled. If an unexpected state is decoded at any given step, it will lead to a deviation of the drum density of future steps too. To reject such samples, we calculate the density of each bar in the 11-bar drum sample and put a cap of 5
% % (values selected by considering the distribution over the dataset)
% on its standard deviation.

We screen the generated samples to eliminate those with more than 4 silent bars and those where the variation of bar density is high as measured by the standard deviation of the bar density. After applying the mentioned filtering to 8192 P1 drum samples generated by simple sampling method, we are left with 3762 samples for further evaluation.
% The above filtration methods are applied on $\approx 8000$ generated drum samples and we are left with 3762 samples.

% \subsection{Objective Evaluation}\label{subsec:evaluation_objective}

% Evaluation of generative music systems faces harder challenges than that of image generation systems (Briot et al. \cite{briot2017deep}).

% A basic requirement that any drum pattern generation system needs is the rhythmic positioning of onsets. Within a bar, the first beat (downbeat) along with beats 2 and 4 (backbeat) are emphasized in rock music, wherein the downbeat is generally identified by a crash cymbal, and the backbeat is associated with a snare drum hit. \figref{fig:result_combined}a shows the onset locations distribution present in the bar of both O and P1 samples. We find a larger proportion of onsets at the accented locations within a beat interval.

\begin{figure}[ht]
    \centering
    \includegraphics[width=\linewidth]{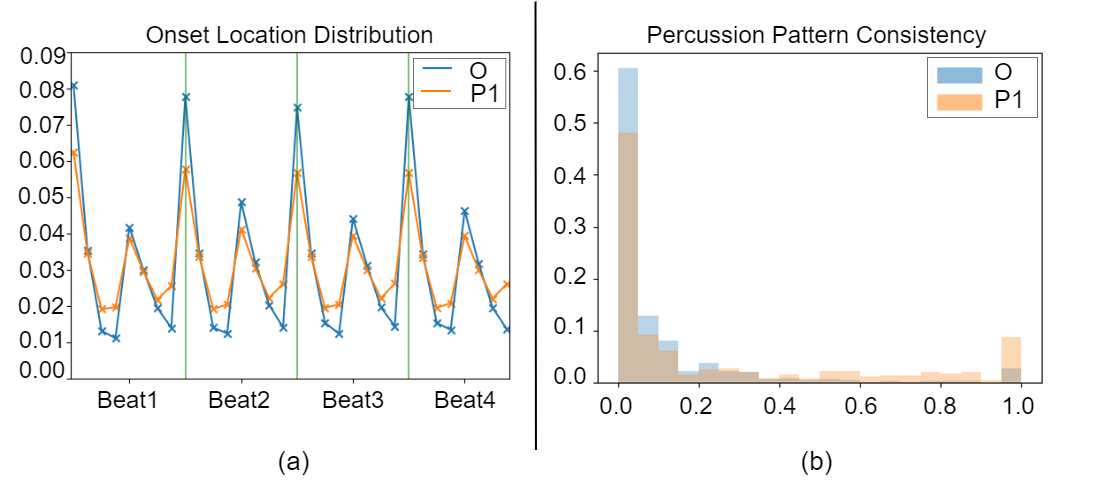}
    \vspace{-20pt}
    \caption{(a) Distribution of onset position in a bar (b) Distribution of Percussion Pattern Consistency}
    \label{fig:result_combined}
    \vspace{-5pt}
\end{figure}

A basic requirement that any drum pattern generation system must fulfill is the rhythmic positioning of onsets. Generally, in a bar of rock music, the first beat (downbeat) and the third beat have similar instrumentation, featuring a hi-hat and kick drum onset. Beats 2 and 4, commonly referred to as the backbeat, include a hi-hat and snare drum onset. While the quarter notes are accented across the bar, $8^{th}$ notes are accented within the beat interval. \figref{fig:result_combined}a shows the onset distribution present in the bar of both O and P1 samples. We can observe that most of the drum patterns are $8^{th}$ note patterns and we find a larger proportion of onsets at the accented locations within a beat interval. 
% This is particularly interesting as no explicit downbeat and sub-beat information is given to the model, yet it is able to accent the sub-beats required for 8 beat pattern.
This is particularly intriguing because the model is given no explicit downbeat or subdivision information yet is still able to emphasis the subdivisions required for an $8^{th}$ note pattern.

We also do a one-to-one comparison of the P1 outputs against their target drum pattern to understand how closely the patterns match with the original drum sample based on the following metric:\\
\textbf{Instrument Count} (IC) is defined as the total number of distinct instruments used in a bar. To see whether our model is able to replicate the behavior of multi-instrument dependency, we calculate the deviation of IC in the generated sample with reference to the original target drum track for the same MA. We observe that in 75.8\% of the drum bars, we are able to replicate the IC, while in 99.3\% of the bars our model was off by at most 1 instrument.

% using the following metrics:

% \begin{figure}[ht]
%     \centering
%     \includegraphics[width=\linewidth]{figures/result_diversity.png}
%     \caption{}
%     \label{fig:result_id_sd}
% \end{figure}

% 1. \textbf{Instrument Count (IC):} Instrument count is defined as the total number of distinct instruments used in a bar. To get a better understanding of whether our model is able to replicate the behavior of multi-instrument dependency, we calculate the error of IC in the generated sample with that present in the real drum track for the same MA. We observe that in about 75.8\% of the drum bars, we are able to replicate the IC, while in 99.3\% of the bars our model was off by at most 1 instrument.
    
% 2. \textbf{Stroke Count (SC):} Stroke count is defined as the total number of timesteps with the onset of any percussion instruments. To understand how well our model was able to replicate the net strokes in a bar, we calculate the error of SC in the generated sample with that present in the real drum track for the same MA. \figref{fig:result_combined}b shows the distribution of this error. We can see that while most of the time the model was able to generate similar stroke count as the ones present in the dataset, some bars had a significantly larger error. The model compensated an increase in SC for one bar by slightly lowering the SC at some other bar, on an average.

Another important aspect that needs to be considered while generating drum patterns is to have a rhythmic (pattern) consistency across bars. %This means that the generated drum bars should have a consistent drum pattern.
% and there needs to be a consistency in the onset locations across the bars. 
We evaluate this aspects of the generated drum pattern using the following metric:\\
\textbf{Pattern Consistency:} For consecutive bar pair, we calculate the distance between the drum patterns using \eqref{equ:distance_formula} keeping $k=1$. The distribution of the bar distances is shown in \figref{fig:result_combined}b. We can see that the generated drum bars are more or less similar to each other with some minor deviations due to the sampling decoding method. The overlapping area of the two distributions is 80.4\%.

% 1. \textbf{Inter-onset interval (IOI):} Inter-onset interval in the symbolic domain is the number of timesteps between two consecutive onsets. We measure the IOI on the dataset and the generated samples and compute its distribution and find that the overlapping area of the two is 93.2\%.
    
% 2. \textbf{Pattern Consistency:} For consecutive bar pair, we calculate the distance between the drum patterns using \eqref{equ:distance_formula} keeping $k=1$. The distribution of the bar distances is shown in figure \ref{}. We can see that the generated drum bars are more or less similar to each other with some minor deviations due to the sampling decoding method. The overlapping area of the two distributions is 80.4\%.

% To understand how good were our models able to capture the fills/improvisations, we use the following evaluation methods on the improvised O and P2 bars:

\begin{figure}[t]
    \centering
    \includegraphics[width=\linewidth]{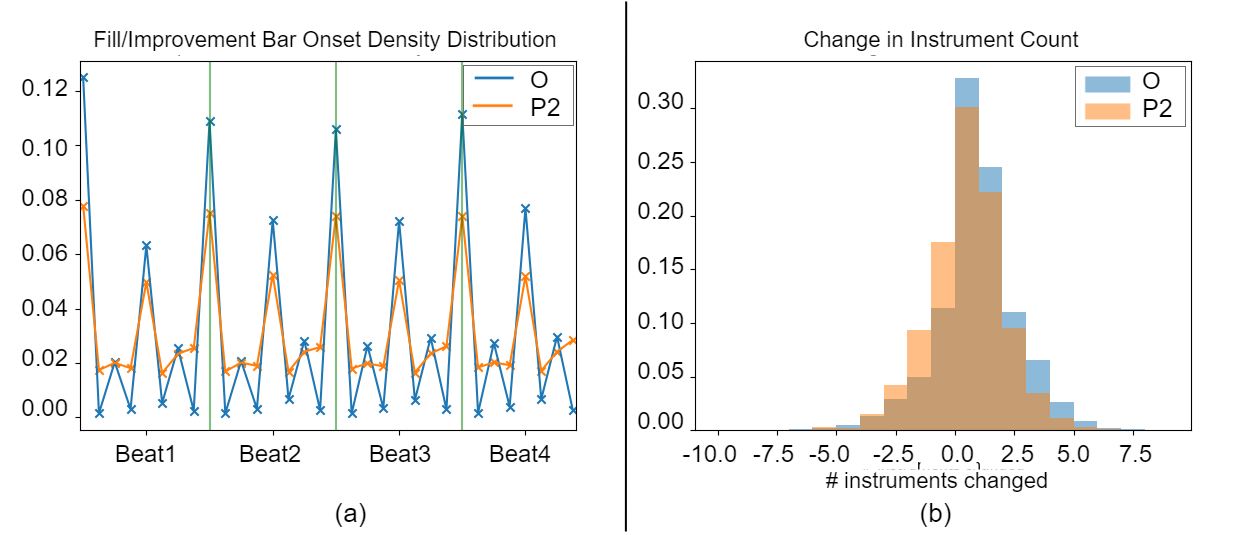}
    \caption{Improvised bars: Distribution of (a) onset location (b) change in instrument count (w.r.t. previous bar)}
    \label{fig:o_p2_comparison}
    \vspace{-9pt}
\end{figure}

Next, on the improvised O and P2 bars, we used the following evaluation methods to see how well our models captured the fills/improvisations:\\
\textbf{Onset Position:} \figref{fig:o_p2_comparison}a shows the distribution of onset location of percussion instruments across the improvised bars. We observe a slightly higher proportion of $16^{th}$ note patterns in the improvised O bars as compared to onset distribution across the non-improvised bars seen in \figref{fig:result_combined}a. We can see that P2 system is largely able to capture this behavior as well.\\
\textbf{Instrument Count (IC) Change:} Generally during a fill/improvisation, some additional instrument are being introduced. IC change is measures as the change in IC of improvised bar compared to its previous bar. \figref{fig:o_p2_comparison}b shows the distribution of IC change. The overlapping area of the two distributions is 87.9\%, This shows that our model was able to capture the general trend of IC change.

%%%%%%%%% SUBJECTIVE EVALUATION %%%%%%%%%
\changes{
% RD: These 2 paragraphs instead of the complete subjective evaluation section.

Additionally, to evaluate the perceptual quality of the generated outputs, we present the generated samples to trained musicians.
% 7 MA tracks and their corresponding PA from the 3762 filtered samples were selected.
We created 3 pairs i.e. O \& P1; P1 \& P2; O \& P2 for each MA-PA track and presented them to two guitarists and a multi-instrumentalist with experience ranging from 5 to 10 years. They were asked to provide detailed comments on the drum pattern in terms of timing, appropriateness of fills and coherence of the PA with MA. %During the comparison, they were asked not to focus on the audio quality or loudness.
A common comment from the musicians was regarding the monotonicity in P1 track. As a result when the O \& P1 pair was presented, majority of the times O was preferred, but when P1 \& P2 were presented, musicians were found to appreciate the fills as it provided a lively feel to the PA.

% In terms of coherence of melody and the generated drum track, while the majority of the time O drum track was preferred over P1, P2 was able to gain significantly more votes over O as compared to P1 over O. The basic drum pattern with no fills/improvisation in P1 gave it a monotonous feel and it was one of the major reason for selecting O over P1.
% % One of the most common reason for selecting O over P1 was the consistent drum pattern with no fills/improvisations in P1 gave it monotonous feel.
% While comparing P1 and P2, in a few samples it was difficult to choose a better version due to difference of just one bar. Most of the times, P2 was selected as the fills imparted a lively feel to the accompanying drums. However, in a few cases, it was also reported that the fill introduced by P2 felt a bit unnecessary.
}

\vspace{-5pt}
\section{Conclusion and Future Work}

We have successfully shown a method to produce coherent drums accompaniment with improvised bars by conditioning on a given melodic accompaniment.  A novel BERT inspired infilling architecture is proposed, along with self-supervised improvisation locator. By learning, where and how to improvise, our evaluations indicate improved generation quality. Thus with a two step approach, we mitigate the biases intrinsic with data-imbalance, and shortcomings that exists with current machine learning architectures. The system can further be improved by learning optimal sampling techniques, which still remains an open problem. As a future work, we could improve the detection performance by employing larger and deeper architectures. %The findings of our work, can yield to advances in other domains such as natural language processing and computer vision due to ubiquity of these architectures. Finally, extending these beyond midi like events and solving this at scale remains exciting and challenging problem ahead of us.
This work highlights a serious drawback of traditional language-based generators, which have shown promise in a lot of different fields, yet they fail to capture subtle musical signals, where they are often sparsely occurring in otherwise repetitive and common patterns.

% For bibtex users:
\nocite{yang2020evaluation}
\bibliography{ISMIRtemplate}

\begin{thebibliography}{10}
\providecommand{\url}[1]{#1}
\csname url@samestyle\endcsname
\providecommand{\newblock}{\relax}
\providecommand{\bibinfo}[2]{#2}
\providecommand{\BIBentrySTDinterwordspacing}{\spaceskip=0pt\relax}
\providecommand{\BIBentryALTinterwordstretchfactor}{4}
\providecommand{\BIBentryALTinterwordspacing}{\spaceskip=\fontdimen2\font plus
\BIBentryALTinterwordstretchfactor\fontdimen3\font minus
  \fontdimen4\font\relax}
\providecommand{\BIBforeignlanguage}[2]{{%
\expandafter\ifx\csname l@#1\endcsname\relax
\typeout{** WARNING: IEEEtran.bst: No hyphenation pattern has been}%
\typeout{** loaded for the language `#1'. Using the pattern for}%
\typeout{** the default language instead.}%
\else
\language=\csname l@#1\endcsname
\fi
#2}}
\providecommand{\BIBdecl}{\relax}
\BIBdecl

\bibitem{vaswani2017attention}
A.~Vaswani, N.~Shazeer, N.~Parmar, J.~Uszkoreit, L.~Jones, A.~N. Gomez,
  {\L}.~Kaiser, and I.~Polosukhin, ``Attention is all you need,''
  \emph{Advances in neural information processing systems}, vol.~30, 2017.

\bibitem{verma2021generative}
P.~Verma and C.~Chafe, ``A generative model for raw audio using transformer
  architectures,'' \emph{arXiv preprint arXiv:2106.16036}, 2021.

\bibitem{brown2020language}
T.~Brown, B.~Mann, N.~Ryder, M.~Subbiah, J.~D. Kaplan, P.~Dhariwal,
  A.~Neelakantan, P.~Shyam, G.~Sastry, A.~Askell \emph{et~al.}, ``Language
  models are few-shot learners,'' \emph{Advances in neural information
  processing systems}, vol.~33, pp. 1877--1901, 2020.

\bibitem{verma2021audio}
P.~Verma and J.~Berger, ``Audio transformers: Transformer architectures for
  large scale audio understanding. adieu convolutions,'' \emph{arXiv preprint
  arXiv:2105.00335}, 2021.

\bibitem{dhariwal2020jukebox}
P.~Dhariwal, H.~Jun, C.~Payne, J.~W. Kim, A.~Radford, and I.~Sutskever,
  ``Jukebox: A generative model for music,'' \emph{arXiv preprint
  arXiv:2005.00341}, 2020.

\bibitem{huang2018music}
C.-Z.~A. Huang, A.~Vaswani, J.~Uszkoreit, N.~Shazeer, I.~Simon, C.~Hawthorne,
  A.~M. Dai, M.~D. Hoffman, M.~Dinculescu, and D.~Eck, ``Music transformer,''
  \emph{arXiv preprint arXiv:1809.04281}, 2018.

\bibitem{holtzman2019curious}
A.~Holtzman, J.~Buys, L.~Du, M.~Forbes, and Y.~Choi, ``The curious case of
  neural text degeneration,'' \emph{arXiv preprint arXiv:1904.09751}, 2019.

\bibitem{ficler2017controlling}
J.~Ficler and Y.~Goldberg, ``Controlling linguistic style aspects in neural
  language generation,'' \emph{arXiv preprint arXiv:1707.02633}, 2017.

\bibitem{fan2018hierarchical}
A.~Fan, M.~Lewis, and Y.~Dauphin, ``Hierarchical neural story generation,''
  \emph{arXiv preprint arXiv:1805.04833}, 2018.

\bibitem{schick2020rare}
T.~Schick and H.~Sch{\"u}tze, ``Rare words: A major problem for contextualized
  embeddings and how to fix it by attentive mimicking,'' in \emph{Proceedings
  of the AAAI Conference on Artificial Intelligence}, vol.~34, no.~05, 2020,
  pp. 8766--8774.

\bibitem{sheng2019woman}
E.~Sheng, K.-W. Chang, P.~Natarajan, and N.~Peng, ``The woman worked as a
  babysitter: On biases in language generation,'' \emph{arXiv preprint
  arXiv:1909.01326}, 2019.

\bibitem{wei2019generating}
I.-C. Wei, C.-W. Wu, and L.~Su, ``Generating structured drum pattern using
  variational autoencoder and self-similarity matrix.'' in \emph{ISMIR}, 2019,
  pp. 847--854.

\bibitem{tamagnan2019drum}
F.~Tamagnan and Y.-H. Yang, ``Drum fills detection and generation,'' in
  \emph{International Symposium on Computer Music Multidisciplinary
  Research}.\hskip 1em plus 0.5em minus 0.4em\relax Springer, 2019, pp. 91--99.

\bibitem{dong2018musegan}
H.-W. Dong, W.-Y. Hsiao, L.-C. Yang, and Y.-H. Yang, ``Musegan: Multi-track
  sequential generative adversarial networks for symbolic music generation and
  accompaniment,'' in \emph{Proceedings of the AAAI Conference on Artificial
  Intelligence}, vol.~32, no.~1, 2018.

\bibitem{huang2020pop}
Y.-S. Huang and Y.-H. Yang, ``Pop music transformer: Beat-based modeling and
  generation of expressive pop piano compositions,'' in \emph{Proceedings of
  the 28th ACM International Conference on Multimedia}, 2020, pp. 1180--1188.

\bibitem{ren2020popmag}
Y.~Ren, J.~He, X.~Tan, T.~Qin, Z.~Zhao, and T.-Y. Liu, ``Popmag: Pop music
  accompaniment generation,'' in \emph{Proceedings of the 28th ACM
  International Conference on Multimedia}, 2020, pp. 1198--1206.

\bibitem{nuttall2021transformer}
T.~Nuttall, B.~Haki, and S.~Jorda, ``Transformer neural networks for automated
  rhythm generation,'' 2021.

\bibitem{thorn2020ai}
O.~Th{\"o}rn, ``Ai drummer-using learning to enhancearti cial drummer
  creativity,'' 2020.

\bibitem{caliskan2017semantics}
A.~Caliskan, J.~J. Bryson, and A.~Narayanan, ``Semantics derived automatically
  from language corpora contain human-like biases,'' \emph{Science}, vol. 356,
  no. 6334, pp. 183--186, 2017.

\bibitem{huang2019reducing}
P.-S. Huang, H.~Zhang, R.~Jiang, R.~Stanforth, J.~Welbl, J.~Rae, V.~Maini,
  D.~Yogatama, and P.~Kohli, ``Reducing sentiment bias in language models via
  counterfactual evaluation,'' \emph{arXiv preprint arXiv:1911.03064}, 2019.

\bibitem{raffel2016learning}
C.~Raffel, \emph{Learning-based methods for comparing sequences, with
  applications to audio-to-midi alignment and matching}.\hskip 1em plus 0.5em
  minus 0.4em\relax Columbia University, 2016.

\bibitem{briot2017deep}
J.-P. Briot, G.~Hadjeres, and F.-D. Pachet, ``Deep learning techniques for
  music generation--a survey,'' \emph{arXiv preprint arXiv:1709.01620}, 2017.

\bibitem{kingma2014adam}
D.~P. Kingma and J.~Ba, ``Adam: A method for stochastic optimization,''
  \emph{arXiv preprint arXiv:1412.6980}, 2014.

\bibitem{abadi2016tensorflow}
M.~Abadi, P.~Barham, J.~Chen, Z.~Chen, A.~Davis, J.~Dean, M.~Devin,
  S.~Ghemawat, G.~Irving, M.~Isard \emph{et~al.}, ``Tensorflow: A system for
  large-scale machine learning,'' in \emph{12th $\{$USENIX$\}$ symposium on
  operating systems design and implementation ($\{$OSDI$\}$ 16)}, 2016, pp.
  265--283.

\bibitem{agarap2018deep}
A.~F. Agarap, ``Deep learning using rectified linear units (relu),''
  \emph{arXiv preprint arXiv:1803.08375}, 2018.

\bibitem{gillick2021drumroll}
J.~Gillick, J.~Yang, C.-E. Cella, and D.~Bamman, ``Drumroll please: Modeling
  multi-scale rhythmic gestures with flexible grids,'' \emph{Transactions of
  the International Society for Music Information Retrieval}, vol.~4, no.~1,
  2021.

\bibitem{liumulti}
G.-H. Liu, A.~Siravuru, S.~Prabhakar, M.~Veloso, and G.~Kantor, ``Multi-modal
  deep reinforcement learning with a novel sensor-based dropout.''

\bibitem{devlin2018bert}
J.~Devlin, M.-W. Chang, K.~Lee, and K.~Toutanova, ``Bert: Pre-training of deep
  bidirectional transformers for language understanding,'' \emph{arXiv preprint
  arXiv:1810.04805}, 2018.

\bibitem{huber1992robust}
P.~J. Huber, ``Robust estimation of a location parameter,'' in
  \emph{Breakthroughs in statistics}.\hskip 1em plus 0.5em minus 0.4em\relax
  Springer, 1992, pp. 492--518.

\bibitem{pepino2021emotion}
L.~Pepino, P.~Riera, and L.~Ferrer, ``Emotion recognition from speech using
  wav2vec 2.0 embeddings,'' \emph{arXiv preprint arXiv:2104.03502}, 2021.

\bibitem{tolstikhin2021mlp}
I.~O. Tolstikhin, N.~Houlsby, A.~Kolesnikov, L.~Beyer, X.~Zhai, T.~Unterthiner,
  J.~Yung, A.~Steiner, D.~Keysers, J.~Uszkoreit \emph{et~al.}, ``Mlp-mixer: An
  all-mlp architecture for vision,'' \emph{Advances in Neural Information
  Processing Systems}, vol.~34, 2021.

\bibitem{gillick2019learning}
J.~Gillick, A.~Roberts, J.~Engel, D.~Eck, and D.~Bamman, ``Learning to groove
  with inverse sequence transformations,'' in \emph{International Conference on
  Machine Learning}.\hskip 1em plus 0.5em minus 0.4em\relax PMLR, 2019, pp.
  2269--2279.

\bibitem{yang2020evaluation}
L.-C. Yang and A.~Lerch, ``On the evaluation of generative models in music,''
  \emph{Neural Computing and Applications}, vol.~32, no.~9, pp. 4773--4784,
  2020.

\end{thebibliography}

% For non bibtex users:
%\begin{thebibliography}{citations}
% \bibitem{Author:17}
% E.~Author and B.~Authour, ``The title of the conference paper,'' in {\em Proc.
% of the Int. Society for Music Information Retrieval Conf.}, (Suzhou, China),
% pp.~111--117, 2017.
%
% \bibitem{Someone:10}
% A.~Someone, B.~Someone, and C.~Someone, ``The title of the journal paper,''
%  {\em Journal of New Music Research}, vol.~A, pp.~111--222, September 2010.
%
% \bibitem{Person:20}
% O.~Person, {\em Title of the Book}.
% \newblock Montr\'{e}al, Canada: McGill-Queen's University Press, 2021.
%
% \bibitem{Person:09}
% F.~Person and S.~Person, ``Title of a chapter this book,'' in {\em A Book
% Containing Delightful Chapters} (A.~G. Editor, ed.), pp.~58--102, Tokyo,
% Japan: The Publisher, 2009.
%
%
%\end{thebibliography}

\end{document}